\documentclass[aps,twocolumn,prl,nobibnotes,nofootinbib,showpacs]{revtex4}%
\usepackage{graphicx}
\usepackage{amssymb}
\usepackage{color}
\usepackage{amsmath}
\usepackage{amsfonts}%
\begin{document}
\title{Electric-field induced dipole blockade with Rydberg atoms}
\author{Thibault Vogt}
\author{Matthieu Viteau, Amodsen Chotia}
\author{Jianming Zhao\thanks{Visitor from College of Physics and Electronics
Engineering, Shanxi University, China. }}
\author{Daniel Comparat}
\author{Pierre Pillet}
\affiliation{Laboratoire Aim\'{e} Cotton,\footnote{Laboratoire Aim\'{e} Cotton is
associated to Universit\'{e} Paris-Sud and belongs to F\'{e}d\'{e}ration de
Recherche Lumi\`{e}re Mati\`{e}re (LUMAT).} CNRS, B\^{a}t. 505, Campus
d'Orsay, 91405 Orsay, France}
\date\today

\begin{abstract}
High resolution laser Stark excitation of $np$  ($60<n<85$) Rydberg states of ultra-cold
cesium atoms shows an efficient blockade of the excitation attributed to long-range dipole-dipole interaction. The dipole blockade effect is observed as a quenching of the Rydberg excitation  depending on the value of the dipole moment induced by the external electric field.
Effects of eventual ions which could match the dipole blockade effect are discussed in detail but are ruled out for our experimental conditions. 
Analytic and Monte-Carlo simulations of the excitation of an ensemble of interacting Rydberg atoms agree with the
experiments indicates a major role of the nearest neighboring Rydberg atom. 

\end{abstract}
\maketitle

Long-range dipole-dipole interactions often play an important role in the
properties of an assembly of cold atoms. 
One example is the efficiency of photoassociation of cold atoms and the formation of cold molecules \cite{FiorettiA}. 
In the case of a Rydberg atomic ensemble, the range of the dipole-dipole interactions can exceed several micrometers, leading to many-body effects \cite{Mourachko,Anderson,Fioretti}.
 An interesting application of the dipole-dipole interaction is the dipole
blockade (DB) in Rydberg excitation. This effect offers exciting possibilities for
quantum information \cite{Jaksch} with the fascinating possibilities for
manipulating quantum bits stored in a single collective excitation in mesoscopic
ensembles, or for realizing scalable quantum logic gates \cite{Lukin}. The DB
process for an ensemble of atoms is the result of shifting the Rydberg energy
from its isolated atomic value due to the dipole-dipole interaction with the surrounding atoms. In a large volume, a partial or local blockade, corresponding
to a limitation of the excitation is expected when the  dipole-dipole  energy shift 
exceeds   the resolution of the laser excitation.
In a zero electric field,  Rydberg atoms present no permanent dipole and usually no DB is
expected. Nevertheless, a van der Waals blockade, corresponding to a second order
dipole-dipole interaction, has been observed through a
limitation of the excitation of high Rydberg states $np$
($n\sim70-80$) of rubidium, using a pulsed amplified single mode laser
\cite{Tong}. 
CW excitations have also been performed \cite{RaithelPRL05,Singer} showing the suppression of the
excitation  and affecting the atom counting statistics \cite{RaithelPRL05}. 
The DB phenomenon itself has been
observed for the first time, in the case of cesium
Rydberg atoms,
 for a so called F\"{o}rster Resonance
Energy Transfer (FRET) reaction, $np+np\longrightarrow ns+(n+1)s$ \cite{VogtPRL2006}.
The FRET configuration has several advantages: the dipole-dipole interaction can be tuned on and off by the Stark effect,  the dipole-dipole interaction  having its maximum effect at the resonant field. Its main drawback comes from the fact that
the  resonance exists only for $n\leq41$ in the cesium case, which limited the observed DB to an efficiency of $\sim30\%$.

\begin{figure}[ht]
\begin{flushleft}
\resizebox{0.45\textwidth}{!}{
\rotatebox{-90}{\includegraphics*[8mm,9mm][107mm,237mm]{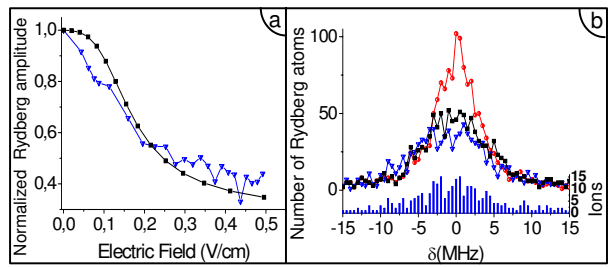}}}
\caption{(Color online) Comparison of the DB with the ion blockade effect. 
Number of atoms excited from ground state to the $70 p_{3/2}$ Rydberg state after a 300ns laser excitation.
(a) Reduced density matrix model (see text) for different external electric field values.  Down triangles: a single ion is randomly present in the sample before the laser excitation but no dipole-dipole interaction is taken into account. Squares:  calculation taking into account only the dipole-dipole interaction with no ions .
(b) Kinetic Monte Carlo simulation of the laser excitation, versus the detuning from resonance, of $8500$ ground state atoms, in a spherical gaussian cloud of $\sigma =30\,\mu$m ($1/e^{1/2}$) radius at a given electric field (F=0.4V/cm). 
No dipole interactions (circles), with no dipole interactions but with ions appearing from Rydberg ionization (down triangles), dipole interactions but no ions (squares).}
\label{fig:simu}
\end{flushleft}
\end{figure}

In this letter,
we report the experimental evidence for a DB with an efficiency
larger than 60 \%. In order to obtain higher couplings than the van der Waals or F\"{o}rster ones, we apply an electric field to create a significant dipole moment.
Indeed, the Stark Rydberg state, $np$, presents an
electric dipole moment mostly due to their mixing with the $(n-1)d$ state. It is however experimentally challenging to observe  the dipole blockade effect without any ambiguity. Ions could for instance lead to a similar effect as the DB one as  shown by Fig \ref{fig:simu} [discussed latter]. The letter is presented as following.
First we discuss the role of ions and the importance of removing almost all ions in the experiment. We then report the DB results which are finally compared to an analytical model. 

In every Rydberg DB experiment the role of ions needs to be studied carefully because
the  presence of a single ion creates a spurious electric field of $1.5\,$mV/cm at a distance of $100\,\mu$m.
The role of ions is often underestimated and can lead to a blockade similar to the Van der Waals blockade effect.
Furthermore, due to the quadratic variation of the Rydberg $np$ energy level as a function of the external electric field,  the Stark effect created by ions is enhanced compared to the zero field configuration.
In  the F\"{o}rster configuration, due to the  low $n$ value,
 it was clearly seen \cite{VogtPRL2006} that the presence of some ions did not affect the observed limitation of the excitation.  However,
 for higher $n$ value, the  presence of a single ion
 during
the excitation 
could easily, even for a moderate external electric field,  shifts the Rydberg energy level by few
 MHz, which is 
on the order of  the laser
resolution. In other
words, the appearance of a single ion can stop the
excitation. 
To investigate in more detail how strong the effect of ions is,
two kinds of numerical simulations have been performed. The result of the first model (see Fig.\ref{fig:simu} (a)), based on single atoms reduced density matrix evolution in the presence of the nearest neighbors dipole-dipole interaction (detailed hereafter), 
illustrates 
the role of a single ion present at the beginning of the excitation towards the 70p state. Such ion would affect drastically the results because it has almost the same effect as the dipole-dipole interaction without any ion present. 
However, in our experiment the ions do not appear at the beginning of the excitation, but only when Rydberg atoms are present. 
Indeed,
we have experimentally studied
the temporal evolution of the number of ions and found a constant rate of ionization and a linear dependence on the number of Rydberg atoms.
The origins of the ions are mainly thermal blackbody radiation and collisional processes between Rydberg atoms and surrounding hot $6s$ or cold $6p,7s$ atoms (in preparation). Rydberg atom-Rydberg atom collisions, such as
Penning ionization occurring from the pair dynamics under the influence of the attractive long-range forces \cite{2005PhRvL..94q3001L,2007PhRvL..98b3004A},
occur only after one microsecond. 
We have then derived a second model taking into account a constant ionization rate of the Rydberg atoms. 
The result of the second model, based on a kinetic Monte Carlo simulation, 
illustrates (see Fig.\ref{fig:simu} (b)) 
the role of such a  Rydberg ionization, but here with a rate 20 times larger than the experimental one, if present during the excitation. This would also affect drastically the results because few ions have almost the same effect than the dipole-dipole interaction when no ions are present. 
The kinetic Monte-Carlo algorithm has been choosen because it gives the exact time evolution of  rate equations system \cite{2003cond.mat..3028J}. Here the rate equations describe  the (2-level) atoms excitation, the sum of all possibles two-body dipole-dipole interactions and the external and ionic electric fields effects. 
In conclusion, the main result of these two types of simulations is that ions can
lift any Rydberg density effect such as the DB one.
 
 Therefore,
the number of ions has to be minimized experimentally, for instance by avoiding the excitation of too many Rydberg atoms. The details of the experimental setup have been
described in reference \cite{VogtPRL2006}. The Rydberg atoms are
excited from a cloud of 5$ \times10^{7}$ cesium atoms (characteristic radius $\sim$ 300 $\mu$m, peak density $1.2\times10^{11}%
$ cm$^{-3}$) produced in a standard vapor-loaded magneto-optical trap (MOT). The first step of the excitation, $6s,F=4\rightarrow6p_{3/2},F=5$, is provided either
by the trapping lasers (wavelength: $\lambda_{1}=852$ nm) or by an independent laser. The density of
excited, $6p_{3/2}$, atoms can be modified by switching off the repumping
lasers before the excitation sequence. The second step, $6p_{3/2}%
,F=5\rightarrow7s,F=4$, is provided  by an infrared diode laser in an extended
cavity device from TOPTICA (wavelength: $\lambda_{2}=1.47$ $\mu$m, bandwidth: 100 kHz). The experimental average intensity is $\sim3$
mW/cm$^{2}$ twice the saturation one. The last step of the excitation,
$7s,F=4\rightarrow np_{1/2,3/2}$ (with $n=25-140$), is provided by a
Titanium:Sapphire laser (Ti:Sa), wavelength $\lambda_{3}$=$770-800$ nm, bandwidth: $1$ MHz. The Ti:Sa laser is switched on with a fixed optical frequency during a time,
$\tau=0.3\,\mu$s, by means of an acousto-optic modulator.   Due to the short $7s$ lifetime and the short excitation time  the spectral resolution of the excitation $\Delta_{L}$ is only on the order of
$5$ MHz (when no Ti:Sa optical saturation is present).
The
Rydberg atoms are selectively field ionized by applying, at a chosen time (between $0$ and $1\,\mu$s) after the end of the Ti:Sa laser pulse, a high field pulse with a rise time of 700 ns. We  use a time of flight technique and a Micro Channel Plate (MCP) detector to know the total number of detected ions at the MCP. From the $\sim 30\%$ quantum detection efficiency of the MCP we estimate the number of Rydberg atoms present in the experimental area. We define the density as $n={N}/(2\pi \sigma^{2})^{3/2}$.
 The beams of the
infrared diode laser and Ti:Sa laser are focused into the atomic cloud with a ($1/e^2$) radius $2 \sigma =70-125\,\mu$m. 
The repetition rate of the experiment is 80 Hz.
The Ti:Sa laser polarization is linear and parallel to the
direction of the applied electric field, leading to the excitation of the
magnetic sublevel, $np_{1/2}$ or $np_{3/2}$ $\left\vert m\right\vert =1/2$.
For clarity we will present here only results obtained with $np_{3/2}$ atoms, 
but similar results have been obtained when using the $np_{1/2}$ state \cite{These_Thibault}.

\begin{figure}[ht]
\begin{center}
\resizebox{0.45\textwidth}{!}
{
\includegraphics*[12mm,182mm][200mm,269mm]{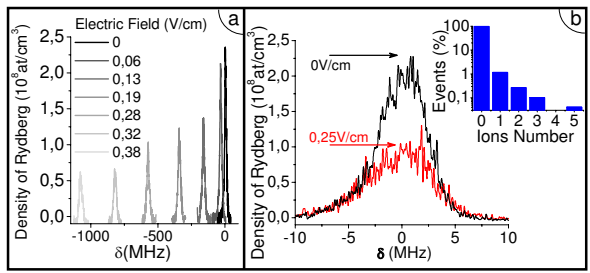}
}
\end{center}
\caption{(Color online)  (a) Excitation spectral lines of the $75p_{3/2}$ level recorded for different electric fields with $\sigma=120\mu m$.
(b) Excitation spectral lines of the $70p_{3/2}$ level for two different fields, as a function of the detuning from the resonance frequency of the transition $7s \rightarrow 70p_{3/2}$ with $\sigma=35\mu m$. Events of MCP detected ions during the scan are also presented (inset, log scale).}%
\label{fig:levels}%
\end{figure}

Fig.\ref{fig:levels} (a) displays an ensemble
of spectra of the excitation of the level, $75p_{3/2}$, for different values
of the static electric field, $F$, indicating the limitation of the excitation induced by the electric field.
Fig.\ref{fig:levels} (b) displays two spectra of the resonance line for the
excitation of the level, $70p_{3/2}$, at zero field  and at $F=0.25$ V/cm.  A clear effect
 appears at the center of the resonance line with a 60\%
decreasing of the intensity of the line. 
At this point, it is very important to know if this reduction of the excitation can be attributed to DB or due to
ion blockade. 
In the present experiment, for evacuating the role of any
ions, we have performed the experiment with conditions where the observed number of ions is on average well below unity as shown in inset of Fig. \ref{fig:levels} (b).
This absence of ions is confirmed by the fact that
 the appearance of ions would be correlated with the Rydberg signal. Indeed, in the simulation, but not in the experiment, we observe a correlation between the ions and the Rydberg Signal, leading to the limitation or blockade of the excitation when ions appear (see Fig \ref{fig:simu} (b), down triangles). 

We then attribute the limitation of the excitation
to a DB and not to an ion blockade.
We stress here that the wings of the resonance should (and are) not affected 
by the DB because the number of excited Rydberg atoms is reduced.  However, in our experiment an apparent increasing of the linewidth (FWHM)
is observed because the line has no more a Lorentzian shape.

\begin{figure}[ht]
\begin{center}
\resizebox{0.43\textwidth}{!}{
\includegraphics*[10mm,177mm][200mm,267mm]{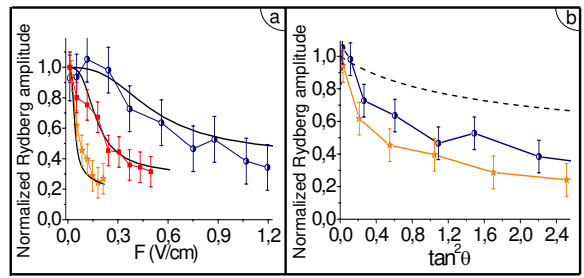}}
\end{center}
\caption{(Color online) (a) Fraction of the number of Rydberg atoms excited at different electric fields compared to the number of Rydberg atoms excited at zero field, for different $np_{3/2}$ states, with $n$ equal to 60 (circles), 70 (squares)and  85 (stars). The Ti\negthickspace:Sa laser intensity is given by $(n/85)^3 \times 560$ W/cm$^2$. Solid lines are given by the reduced density matrix model. (b) Data for $n$=60 and $n$=85 [same symbols as (a)], characterized by $\tan^2 \theta$ (see text), the theoretical curve (dashed line)  is obtained without the DB. }%
\label{fig:DipBlockadeP3/2} 
\end{figure}

In Fig. \ref{fig:levels} (a) we have observed a quite efficient decrease of the number of excited atoms with $F$ for $n=75$ due to the DB effect coming from the existence of a non zero permanent dipole, aligned with the static electric field $\overrightarrow{F}$,
 $ \left\langle np ,F\right\vert q_e
\overrightarrow{r}\left\vert np,F\right\rangle
 = \overrightarrow{\mu}$.
 Similarly, Fig.\ref{fig:DipBlockadeP3/2} (a) demonstrates the DB
effect for the excitation of different levels, $np_{3/2}$, with $n$ equal to 60, 70
and 85. In this study, the intensity of the Ti:Sa laser is set to be inversely
proportional to the oscillator strength of the transition, $7s\rightarrow
np$, meaning a $n^{3}$ dependence. 
To compare the
curves obtained for different principal quantum numbers $n$, it is convenient
to introduce the scale parameter $\theta$ characterizing the dipole coupling
for each level, $np$ defined by %
$
\tan\theta=\frac{\left\vert W_{n}\right\vert }{h \Delta_{n}/2},
$
where $h \Delta_{n}$ is the zero field energy difference between the $(n-1)d$ and $np$
levels and $W_{n}$ the Stark coupling %
$
W_{n}=\left\langle (n-1)d_j, m  \right\vert -q_e \overrightarrow{r}.\overrightarrow
{F}\left\vert np_j m  \right\rangle
$.
Fig. \ref{fig:DipBlockadeP3/2} (b) shows the decreasing of the Rydberg
excitation for $np_{3/2}$ states, with $n$ equal to 60 and 85, versus $\tan^{2}\theta$. We observe a more efficient DB
effect for $n=85$ than for $n=60$ when compared with the calculated (dashed) curve without the dipole-dipole interaction.

To model our data, we have developed a model based on the
interaction of each atom $i$, with its Rydberg neighbors $j$ (more details can  be found in \cite{These_Thibault}). 
If atom $i$ could be  excited to the Rydberg state its energy would be shifted by 
an amount of \[
W_{ij}=\frac{\overrightarrow{\mu}_{i}.\overrightarrow{\mu}_{j}-\left(
\overrightarrow{\mu}_{i}.\overrightarrow{n}_{ij}\right)  \left(
\overrightarrow{\mu}_{j}.\overrightarrow{n}_{ij}\right)  }{4 \pi \epsilon_{0} R_{ij}^{3}}%
\]
where $\overrightarrow{n}_{ij}=\overrightarrow{R}_{ij}/R_{ij}$ with $R_{ij}$ the interatomic distance between the Rydberg atoms $i$ and $j$. 
Due to the $1-3 \cos^2 \psi$ variation of the $W_{ij}$, where
$\psi$ is the angle between the internuclear axis and
the dipoles (i.e. the external field), the angular averaging of $W_{ij}$ is zero.
Therefore, we need to discretized the averaging. We found by the Monte
Carlo simulation that the nearest neighbor interaction dominates. We
then model the dipole-dipole interaction using only the nearest neighbor atom $j$  and
neglect the mean-field interaction. This is the opposite in the van de
Waals interaction where we neglect the nearest neighbor interaction.
$W_{ij}$ is significant only when both atoms are in the Rydberg state. We find 
\[
{Tr_{\neq i}}\left[  W_{ij},\rho\right]  \simeq\frac{\mu^{2}n_{Ry}}{4 \pi \epsilon_{0}}\left(
\overrightarrow{r}_{i}\right)  \left[  \left\vert np\right\rangle
_{ii}\left\langle np\right\vert ,\rho_{i}\right]
\]
which corresponds for the atom $i$ to a shift of the Rydberg level $np$
proportional to the local Rydberg density  $n_{Ry}( \overrightarrow{r_i})$. Here
$\rho$ is the density matrix for the ensemble of the
atoms, and $\rho_{i}={Tr_{\neq i}}\rho$ is the
reduced density matrix for the single atom $i$. 
For a given atom $i$, we solve the evolution equation of the partial density matrix
\[
i\hbar\frac{d\rho_{i}}{dt}=\left[  H_{i},\rho_{i}\right]  +{Tr_{\neq i}}\left[  W_{ij},\rho\right]
\]
$H_{i}$ is the (Bloch eq. type) Hamiltonian describing the 3-step excitation. The spontaneous emission relaxation terms 
 are taken into account but not written here for sake of
simplicity.

\begin{figure}[ht]
\begin{center}
\resizebox{0.45\textwidth}{!}{
\includegraphics*[10mm,193mm][198mm,275mm]{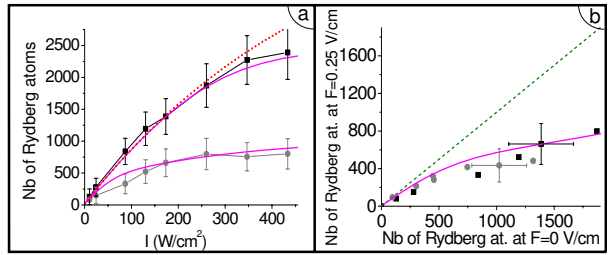}}
\caption{(Color online) Number of Rydberg atoms excited versus the Ti:Sa laser intensity, in the case of the $70p_{3/2}$ state, for (a) $7s$-atom density $D\sim 4 \pm 2 \times 10^9 cm^{-3}$ and in the presence of two different electric fields, 0 V/cm (squares) and $F_1$=0.25 V/cm (circles);  (b) Number of Rydberg atoms excited in the presence of the electric field $F_1$ is plotted versus the number of Rydberg atoms excited at zero field, for two $6p$ $7s$-atom densities, $D$ ( squares) and $D/2.7$ (circles). The DB efficiency is the ratio between by the gap between the experimental points and the straight dashed line of slope 1. In (a) and (b): solid lines are theoretical calculations taking into account the van der Waals  blockade at zero field and the DB in the presence of the electric field $F_1$. In (a), for comparison, we have also presented the curve with no van der Waals coupling (dotted curve).} %
\label{fig:GP32Puissance}%
\end{center}
\end{figure}

Fig. \ref{fig:DipBlockadeP3/2} (a) shows the comparison between the
calculated curves and the experimental ones with a good agreement.
 All curves in this letter, are calculated using the exact dipole moments calculated by considering
the ensemble of the levels of the Stark diagram \cite{Zimmerman}. However,
most of the results can be understood using a 
simpler two-level approach, valid for small electric
fields ($F \ll 1/3n^{5}$, atomic units), 
assuming only mixing of the $np$ state with the $(n-1)d$ one. In this model
the $np$ Rydberg permanent dipole is $\mu\sim q_e a_0n^{2}\sin\theta$ and the excitation (dashed) curve of Fig. \ref{fig:DipBlockadeP3/2} (b) would have a value of $\cos^{2}(\theta/2)$.
The DB 
condition:
$ W_{ij} \sim\frac{\mu^{2}}{4 \pi \epsilon_0 R_{ij}^{3}} < h \Delta_{L}$ can then be simply evaluated.
 This DB condition
gives a limitation to the excitation corresponding to a density
\[
n_{Ry} \sim h\Delta_{L} 4\pi \epsilon_0 /\mu^{2}.
\]
Such a limitation is illustrated by Fig. \ref{fig:GP32Puissance}(a), showing the evolution of the DB versus the Ti:Sa laser intensity in
the case of the 70p$_{3/2}$ Rydberg level, where the DB condition gives N$\sim 800$ Rydberg atoms and fits well to the experiment ($\sigma \sim 50\mu m$).
At the electric field $F_1=0.25$ V/cm (corresponding to $\tan^{2}\theta=1$)
the appearance of the DB
occurs at intensities higher than 100 W/cm$^{2}$.
The DB efficiency is evaluated by comparison with the zero field case.
We
observe a power saturation for intensities larger than 250
W/cm$^{2}$ explaining why
 the maximum DB efficiency is reached  when the intensity is close to 250 W/cm$^{2}$.
 For higher intensity an optical broadening occurs so $\Delta_L$ increases and the DB efficiency decreases.  
 For the zero field case, taking into account the van der Waals coupling between the atoms 
modifies the theoretical curve, giving a
better agreement with the experimental data.
The DB efficiency, evaluated by comparison with the zero field case, is also presented in  Fig.\ref{fig:GP32Puissance}(b) from two data sets taken at two different $6p$ (and so two different $7s$)-atomic densities. 
The DB appears not to be dependent on the $6p,7s$ densities but only on the number of excited Rydberg atoms. This result is intrinsic to the DB and cannot be due to an ion blockade effect coming from Rydberg atoms ionization created by collisions with $6p$ or $7s$ atoms. 


To conclude, we have presented the evidence for an efficient DB controlled via the Stark
effect, of the
Rydberg excitation 
in a cold cesium atomic sample.
The observation of the DB is challenging because no ions should be present
during the short excitation pulse.
An analytical model based on the preferential role of the nearest neighboring Rydberg atoms, confirmed by M.C. simulations,
has been derived and provides a good account of the data.
This result is really promising for quantum information. A
microwave field instead of Stark effect or F\"{o}rster may be used to produce the
dipole moments and to control the DB \cite{PilletPRA87,2006physics..12233B}. 
A particularly promising direction of research is now
the observation of the total DB, meaning the collective
excitation of a single atom. Prospect for quantum gate devices and
control of ultra-cold Rydberg atoms in this quantum regime 
is still a challenge.

This work is in the frame of "Institut francilien de recherche sur les atomes
froids" (IFRAF). The authors thank and acknowledge very fruitful discussions
with Thomas F. Gallagher, Duncan A. Tate, Etienne Brion, Marcel Mudrich, Nassim Zahzam, and
Vladimir Akulin.

\bibliography{Blocade3}

\end{document}